\documentclass[10pt,a4paper,notitlepage]{article}
\usepackage[latin1]{inputenc}
\usepackage{verbatim,graphicx,amsmath,amsfonts,amssymb,mathrsfs}

\def\arraystretch{1.5}

\title{On the Construction of Zero Energy States\\in Supersymmetric Matrix Models IV\footnote{Supported by the Swedish Research Council}}
\author{
	\renewcommand{\thefootnote}{\arabic{footnote}}
	J. Hoppe\footnotemark[1],\quad 
	D. Lundholm\footnotemark[2]
}
\date{\scriptsize{Department of Mathematics\\Royal Institute of Technology, SE-10044 Stockholm, Sweden}}

\begin{document}

\maketitle
\renewcommand{\thefootnote}{\arabic{footnote}}
\footnotetext[1]{hoppe@math.kth.se}
\footnotetext[2]{dogge@math.kth.se}
\renewcommand{\thefootnote}{\arabic{footnote}}

\begin{abstract}
	Simple recursion relations for zero energy states of supersymmetric 
	matrix models are derived by using an unconventional \emph{reducible} 
	representation for the fermionic degrees of freedom.
\end{abstract}

\section{The model}

	The supercharges of the model under consideration,
	\begin{equation} \label{q_def}
		\textstyle
		Q_\beta = \left( -i\partial_{tA} \gamma^t_{\beta\alpha} + \frac{1}{2} f_{ABC}q_{sB}q_{tC} \gamma^{st}_{\beta\alpha} \right) \theta_{\alpha A},
	\end{equation}
	with
	\begin{equation} \label{theta_rels}
		\{ \theta_{\alpha A}, \theta_{\beta B} \} = 2 \delta_{\alpha\beta} \delta_{AB},
	\end{equation}
	satisfy the supersymmetry algebra
	\begin{equation} \label{q_rels}
		\{ Q_{\beta}, Q_{\beta'} \} = \delta_{\beta\beta'} 2H + 4 \gamma^t_{\beta\beta'} q_{tA} J_A,
	\end{equation}
	where
	\begin{equation} \label{h_def}
	\begin{array}{c}
		H = -\Delta + V + \frac{1}{2} W_{\alpha A, \beta B} \theta_{\alpha A} \theta_{\beta B}, \\
		V = - \frac{1}{2} \sum_{s,t=1}^d \textrm{tr}\ [X_s,X_t]^2, \\
		W_{\alpha A, \beta B} = if_{ABC} q_{tC} \gamma^t_{\alpha\beta}
	\end{array}
	\end{equation}
	and
	\begin{equation} \label{j_def}
		\textstyle
		J_A = -i f_{ABC} \left( q_{sB} \partial_{sC} + \frac{1}{4} \theta_{\alpha B} \theta_{\alpha C} \right)
		= L_A + S_A.
	\end{equation}
	As each of the $Q_\beta$ squares to $H$ on gauge-invariant 
	states $\Psi$, i.e. when $J_A \Psi = 0$
	($A = 1,\ldots,N^2-1$ in the case of SU($N$)), 
	it is convenient to sometimes suppress the index $\beta$
	(which, corresponding to the dimensions of the real
	representations for the $\gamma$'s, takes $s_d = 2(d-1)$
	different values iff $d$ = 2, 3, 5 or 9).
	
	Writing
	\begin{equation} \label{p_v_def}
		Q_\beta =: D_{\alpha A} \theta_{\alpha A} 
		= \sum_{a=1}^{2\Lambda} D_a \theta_a,
	\end{equation}
	with $2\Lambda := s_d(N^2-1)$ 
	(the total number of fermionic degrees of freedom),
	and choosing $\gamma^d$ to be diagonal, it immediately follows
	from \eqref{q_rels} and
	\begin{equation} \label{q_d_rels}
		\{ Q_{\beta}, Q_{\beta'} \} = \{ D_a \theta_a, D_b' \theta_b \}
		= [D_a,D_b'] \theta_a \theta_b + D_b'D_a \{ \theta_a, \theta_b \}
	\end{equation}
	that the differential operators $D_a\ (\beta=\beta')$ satisfy
	\begin{equation} \label{d_d_trace}
		D_a D_a = -\Delta + V \pm 2q_{dC} L_C
	\end{equation}
	\begin{equation} \label{d_d_commutator}
		\textstyle
		[D_a,D_b] = W_{ab} \pm 4q_{dC} S_C(a,b), \quad 
		S_C(\alpha A, \beta B) = - \frac{i}{4} f_{ABC} \delta_{\alpha\beta}
	\end{equation}
	and the $\pm$ sign corresponds to
	\begin{displaymath}
		\gamma^d_{\beta\beta} = \left\{ 
		\begin{array}{l}
			+1\ \textrm{for}\ \beta \leq \frac{s_d}{2},\ \textrm{say, and} \\
			-1\ \textrm{for}\ \beta > \frac{s_d}{2}.
		\end{array}
		\right. \phantom{\}}
	\end{displaymath}

\section{Recursive solution in the left-action representation}

	Consider the (reducible) representation of \eqref{theta_rels} in which the $\theta$s
	act by multiplication from the left on the Clifford algebra they generate,
	i.e. on the vector space $\mathscr{P}$ of polynomials
	\begin{equation} \label{theta_algebra}
	\begin{array}{ll}
		\Psi 
		&= \psi + \psi_a \theta_a + \frac{1}{2} \psi_{ab} \theta_a\theta_b + \ldots + \frac{1}{(2\Lambda)!} \psi_{a_1 \ldots a_{2\Lambda}} \theta_{a_1} \ldots \theta_{a_{2\Lambda}} \\
		&= \sum_{k=0}^{2\Lambda} \frac{1}{k!} \psi_{a_1 \ldots a_k} \theta_{a_1} \theta_{a_2} \ldots \theta_{a_k},
	\end{array}
	\end{equation}
	where the coefficients $\psi_{a_1 \ldots a_k}$ are totally antisymmetric in their indices.
	The (graded) Hilbert space of the model, $\mathscr{H} = \oplus_{k=0}^{2\Lambda} \mathscr{H}_k = \mathscr{H}_+ \oplus \mathscr{H}_-$,
	is spanned by such polynomials with $\psi_{a_1 \ldots a_k} \in L^2(\mathbb{R}^{d(N^2-1)})$,
	so that $\Psi$ normalizable corresponds to\footnote{One can define the
	scalar product in $\mathscr{H}$ e.g. by 
	$\langle \Phi, \Psi \rangle = \int \langle \Phi_{\textrm{rev}}^* \Psi \rangle_0$,
	where $(\cdot)_{\textrm{rev}}$ denotes reversion of the order of $\theta$s, 
	$(\cdot)^*$ complex conjugation, and $\langle \cdot \rangle_0$ projection onto 
	grade 0 in $\mathscr{P}$.}
	\begin{equation} \label{psi_norm}
		\int |\psi_{a_1 \ldots a_k}(q)|^2 \prod_{t,A} \textrm{d}q_{tA} < \infty \quad \forall k.
	\end{equation}
	The dimension of this representation ($\dim \mathscr{P} = 2^{2\Lambda}$) is 
	vastly greater than that of the irreducible one, but it is completely 
	reducible -- breaking up block-diagonally into the direct sum of
	$2^\Lambda$ copies of the irreducible one. Hence, any non-trivial solution of 
	$H\Psi = 0$ in $\mathscr{H}$ would imply the existence of a zero-energy
	state in the Hilbert space $\hat{\mathscr{H}}$ corresponding to the
	conventional irreducible representation.

	Letting $Q_\beta$ act on $\mathscr{H}_+$ (the even-grade part of $\mathscr{H}$),
	$Q_\beta \Psi = 0$ amounts to\footnote{Cp. \cite{hoppe} for the corresponding 
	irreducible (but manifest SO($d$)-invariance breaking) formulation.}
	\begin{equation} \label{recursion_rels}
		\textstyle
		D_{[a} \psi_{a_1 \ldots a_{2k}]} = \frac{1}{2k+1} D_c \psi_{aca_1 \ldots a_{2k}},
	\end{equation}
	i.e.
	\begin{equation} \label{recursion_rels_2}
		D_a \psi_{a_1 \ldots a_{2k}} + D_{a_1} \psi_{a_2 \ldots a_{2k} a} + \ldots + D_{a_{2k}} \psi_{aa_1 \ldots a_{2k-1}}
		= D_c \psi_{aca_1 \ldots a_{2k}}.
	\end{equation}
	Acting on \eqref{recursion_rels_2} with $D_a$ and summing over $a$ gives
	$(-\Delta + V \pm 2q_{dC}L_C)\psi_{a_1 \ldots a_{2k}}$ for the first
	term, and $\frac{1}{2}(W_{ac} \pm 4q_{dC}S_C(a,c))\psi_{aca_1 \ldots a_{2k}}$
	on the right hand side.
	What about the $2k$ remaining terms $2k D_aD_{[a_1}\psi_{a_2 \ldots a_{2k}]a}$?
	One has
	\begin{equation} \label{recursion_rels_3}
	\begin{array}{ll}
		D_aD_{a_1}\psi_{a_2 \ldots a_{2k} a} 
		&= D_{a_1}D_a\psi_{a_2 \ldots a_{2k} a} + (W_{aa_1} \pm \ldots)\psi_{a_2 \ldots a_{2k} a} \\
		&= (2k-1) D_{a_1}D_{[a_2}\psi_{a_3 \ldots a_{2k}]} + (W_{aa_1} \pm \ldots)\psi_{a_2 \ldots a_{2k}},
	\end{array}
	\end{equation}
	using $\eqref{recursion_rels}_{k \to k-1}$; so
	\begin{equation} \label{recursion_rels_4}
		2k D_aD_{[a_1}\psi_{a_2 \ldots a_{2k}]a}
		= 2k W_{a[a_1}\psi_{a_2 \ldots a_{2k}]a} \pm \ldots + (2k-1)(2k) D_{[a_1}D_{a_2}\psi_{a_3 \ldots a_{2k}]},
	\end{equation}
	where the last antisymmetrized expression again equals 
	$\frac{1}{2}(W_{[a_1a_2} \pm \ldots)\psi_{a_3 \ldots a_{2k}]}$.
	The terms containing the bosonic $L_A$ and fermionic $S_A$ can either 
	be shown to cancel using the assumption $J_A\Psi=0$, or one simply
	adds the equations resulting for $\beta \leq s_d/2$ to the ones
	resulting for $\beta > s_d/2$.
	In any case, what one can also obtain this way are of course the 
	equations that result by considering $H\Psi = 0$ directly:
	\begin{equation} \label{h_recursion}
	\begin{array}{l}
		(-\Delta + V)\psi_{a_1 \ldots a_{2k}}
		+ 2k W_{a[a_1}\psi_{a_2 \ldots a_{2k}]a}
		+ k(2k-1) W_{[a_1a_2}\psi_{a_3 \ldots a_{2k}]} \\
		\quad = \frac{1}{2} W_{ac}\psi_{aca_1 \ldots a_{2k}}.
	\end{array}
	\end{equation}
	Their recursive solution could proceed as follows:
	The lowest-grade equation
	$(-\Delta + V)\psi = \frac{1}{2}W_{ac}\psi_{ac}$ yields
	\begin{equation} \label{h_recursion_0}
		\textstyle
		\psi = \frac{1}{2}(-\Delta + V)^{-1} W_{ac}\psi_{ac}.
	\end{equation}
	Using \eqref{h_recursion_0} to replace $\psi$ in $\eqref{h_recursion}_{k=1}$,
	$\ldots$, respectively $\psi_{a_3 \ldots a_{2k}}$ in $\eqref{h_recursion}_{k}$
	via the analogue of \eqref{h_recursion_0},
	\begin{equation} \label{h_recursion_k}
		\textstyle
		\psi_{a_3 \ldots a_{2k}} = \frac{1}{2} (H_{2k-2}^{-1} W_{ac} \psi_{ac})_{a_3 \ldots a_{2k}},
	\end{equation}
	\eqref{h_recursion} takes the form
	\begin{equation} \label{h_recursion_compact}
		\textstyle
		(H_{2k}\Psi)_{a_1 \ldots a_{2k}} = \frac{1}{2} W_{ac} \psi_{aca_1 \ldots a_{2k}},
	\end{equation}
	with $H_{2k}$ only acting on $\Psi_{2k} \in \mathscr{H}_{2k}$.
	This procedure is based on the fact that $H_0 = -\Delta + V$
	is invertible and the assumption that this also holds for
	higher-grade $H_{2k}$ on $\mathscr{H}_{2k}$.

\section{Recursion relations in a diagonalizing basis}

	Note that
	\begin{equation} \label{w_eigenvalue_eq}
		\textstyle
		\frac{1}{2} W_{ab}\theta_a\theta_b \left( \psi 
			+ \frac{1}{2} \psi_{a_1a_2}\theta_{a_1}\theta_{a_2} 
			+ \frac{1}{4!} \psi_{a_1a_2a_3a_4}\theta_{a_1}\theta_{a_2}\theta_{a_3}\theta_{a_4} 
			+ \ldots
		\right) \stackrel{!}{=} \mu(q) \Psi
	\end{equation}
	gives the set of equations
	\begin{equation} \label{w_eigenvalue_eqs}
	\begin{array}{l}
		\frac{1}{2} W_{a_2a_1}\psi_{a_1a_2} = \mu\psi \\
		W_{a_1a_2}\psi + W_{a_1a}\psi_{aa_2} - W_{a_2a}\psi_{aa_1} + \frac{1}{2} W_{ab}\psi_{baa_1a_2} = \mu\psi_{a_1a_2} \\
		\vdots
	\end{array}
	\end{equation}
	while $H\Psi \stackrel{!}{=} 0$ in the left-action representation gives
	\begin{equation} \label{h_recursion_eqs}
	\begin{array}{l}
		(-\Delta + V)\psi = \frac{1}{2} W_{ac}\psi_{ac} \\
		(-\Delta + V)\psi_{a_1a_2} + W_{aa_1}\psi_{a_2a} - W_{aa_2}\psi_{a_1a} \\
		\quad +\ W_{a_1a_2} \frac{1}{2} (-\Delta + V)^{-1} W_{ab}\psi_{ab} = \frac{1}{2} W_{ac}\psi_{aca_1a_2} \\
		\vdots
	\end{array}
	\end{equation}
	These equations can be simplified by performing a (pointwise) diagonalization
	$W = UDU^{-1}$, where
	\begin{equation} \label{}
	\begin{array}{c}
		U = [w_1,w_2,\ldots,w_\Lambda, w_1^*,\ldots,w_\Lambda^*], \\
		D = \textrm{diag}(\lambda_1,\ldots,\lambda_{2\Lambda})
		  = \textrm{diag}(\mu_1,\ldots,\mu_\Lambda,-\mu_1,\ldots,-\mu_\Lambda),
		\quad (\mu_k \geq 0).
	\end{array}
	\end{equation}
	
	Corresponding to changing to the space-dependent (non-hermitian) fermion basis
	\begin{equation} \label{theta_tilde}
		\tilde{\theta}_a := (U^\dagger)_{ac} \theta_c = U^*_{ca} \theta_c
	\end{equation}
	which diagonalizes the fermionic part of the hamiltonian,
	\begin{equation} \label{h_f_diag}
		\textstyle
		H_F = \frac{1}{2} W_{ab} \theta_a \theta_b 
		= \frac{1}{2} \sum_c \lambda_c \tilde{\theta}_c^\dagger \tilde{\theta}_c,
	\end{equation}
	one could introduce
	\begin{equation} \label{psi_tilde}
		\tilde{\psi}_{\tilde{a}_1 \ldots \tilde{a}_n} := (U^T)_{\tilde{a}_1a_1} \ldots (U^T)_{\tilde{a}_na_n} \psi_{a_1 \ldots a_n},
	\end{equation}
	i.e. substitute
	\begin{equation} \label{psi_no_tilde}
		\psi_{a_1 \ldots a_n} = (U^*)_{a_1\tilde{a}_1} \ldots (U^*)_{a_n\tilde{a}_n} \tilde{\psi}_{\tilde{a}_1 \ldots \tilde{a}_n}
	\end{equation}
	in all equations, and then use
	\begin{equation} \label{w_diagonal}
		\textstyle
		W_{ab} = \sum_e U_{ae} \lambda_e (U^\dagger)_{eb} = \sum_e U_{ae} \lambda_e U^*_{be}
	\end{equation}
	to simplify the recursion relations.
	Using that
	\begin{equation} \label{}
		\def\arraystretch{1.0}
		U^\dagger U^* =
		\left[
		\begin{array}{cc}
			0 & I \\ I & 0
		\end{array}
		\right]
	\end{equation}
	one finds, e.g.
	\begin{equation} \label{}
		\textstyle
		\frac{1}{2} W_{ac}\psi_{ac\cdots} = \sum_{\underline{e}=1}^\Lambda \mu_{\underline{e}} \tilde{\psi}_{\underline{e},\underline{e}+\Lambda,\cdots}
	\end{equation}
	\begin{equation} \label{}
		\textstyle
		W_{aa_1}\psi_{a_2a} = - \sum_{\tilde{a}_1,\tilde{a}_2} U^*_{a_1\tilde{a}_1} U^*_{a_2\tilde{a}_2} (\lambda_{\tilde{a}_1} \tilde{\psi}_{\tilde{a}_1\tilde{a}_2})
	\end{equation}
	and
	\begin{equation} \label{}
		(H \Psi)_{a_1a_2} = (H)_{a_1a_2,b_1b_2} \psi_{b_1b_2} = U^*_{a_1\tilde{a}_1} U^*_{a_2\tilde{a}_2} (\tilde{H})_{\tilde{a}_1\tilde{a}_2,\tilde{c}_1\tilde{c}_2} \tilde{\psi}_{\tilde{c}_1\tilde{c}_2},
	\end{equation}
	with $\tilde{H}$ being unitarily equivalent to $H$,
	\begin{equation} \label{h_tilde}
		\tilde{H}_{\tilde{a}_1\tilde{a}_2,\tilde{c}_1\tilde{c}_2} := U^T_{\tilde{a}_1e_1} U^T_{\tilde{a}_2e_2} H U^*_{e_1\tilde{c}_1} U^*_{e_2\tilde{c}_2}.
	\end{equation}
	The second equation in \eqref{h_recursion_eqs} thus takes a form
	in which the effective operator on the left hand side becomes
	\begin{equation} \label{h_2_tilde}
	\begin{array}{ll}
		(\tilde{H}_2)_{\tilde{a}_1\tilde{a}_2,\tilde{c}_1\tilde{c}_2}
		=& (\tilde{H}_B)_{\tilde{a}_1\tilde{a}_2,\tilde{c}_1\tilde{c}_2}
			+ (\lambda_{\tilde{a}_2} - \lambda_{\tilde{a}_1}) \delta_{\tilde{a}_1\tilde{c}_1} \delta_{\tilde{a}_2\tilde{c}_2} \\
		& \def\arraystretch{1.0} +\ \left[ \begin{array}{ll} 0&I\\ I&0 \end{array} \right]_{\tilde{a}_1\tilde{a}_2} \lambda_{\tilde{a}_2} \tilde{H}_B^{-1} \left[ \begin{array}{ll} 0&I\\ I&0 \end{array} \right]_{\tilde{c}_1\tilde{c}_2} \lambda_{\tilde{c}_1}.
	\end{array}
	\end{equation}
	Note that
	$(\tilde{H}_B)_{\tilde{a}_1\tilde{a}_2,\tilde{c}_1\tilde{c}_2} = \tilde{T}_{\tilde{a}_1\tilde{a}_2,\tilde{c}_1\tilde{c}_2} + V \delta_{\tilde{a}_1\tilde{c}_1} \delta_{\tilde{a}_2\tilde{c}_2}$
	is unitarily equivalent to 
	$(T+V) \delta_{\tilde{a}_1\tilde{c}_1} \delta_{\tilde{a}_2\tilde{c}_2}$
	(and it may be advantageous to choose a non-canonical representation
	of the momentum operators $p_{tA} = p_a$ in $T = p_ap_a$, to simplify $\tilde{T}$).
	The second term is the analogue of the $\lambda \partial_\lambda$-part of
	the corresponding
	$H_0$ in the space-independent fermions approach (see e.g. \cite{hoppe}), 
	while the third term 
	exclusively acts between \emph{particle-hole pairs},
	as $\tilde{\theta}_{\underline{c}+\Lambda} = \tilde{\theta}_{\underline{c}}^\dagger$
	(this feature, including the particle-hole observation, holds also for
	the higher $k$ equations \eqref{h_recursion}).

\section*{Acknowledgements}

	We thank Volker Bach for fruitful discussions, hospitality, and
	collaboration on closely related subjects, while
	one of us (J.H.) would also like to thank Ki-Myeong Lee for kind hospitality.


\begin{thebibliography}{99}
	\bibitem{hoppe}
	J. Hoppe, \textit{On the construction of zero energy states in supersymmetric matrix models I, II, III,} hep-th/9709132, 9709217, 9711033.
\end{thebibliography}
\end{document}